\documentclass[letter]{aa}  
\usepackage{amsmath,amssymb,latexsym} 
\usepackage{graphicx}
\usepackage{txfonts}
\usepackage{lscape}
\usepackage{longtable}
\usepackage{color}
\usepackage{soul}
\usepackage{caption}
\usepackage{natbib}
\bibpunct{(}{)}{;}{a}{}{,} % to follow the A&A style
\usepackage{url}
\usepackage{hyperref}
\hypersetup{colorlinks,linkcolor=red,citecolor=blue,urlcolor=blue}

\newcommand{\ac}{$\alpha-$capture}
\newcommand{\ip}{iron$-$peak}
\newcommand{\nc}{$n-$capture}
\newcommand{\spro}{$s-$process}
\newcommand{\rpro}{$r-$process}
\newcommand{\oddz}{odd$-Z$}

\begin{document}

   \title{Traits for chemical evolution in solar twins}

   \subtitle{Trends of neutron-capture elements with stellar age}

   \author{Paula Jofr\'e
          \inst{1}
          \and
          Holly Jackson\inst{2}
          \and
          Marcelo Tucci Maia\inst{1}
          }

   \institute{N\'ucleo de Astronom\'ia, 
Universidad Diego Portales, Ej\'ercito 441, Santiago de Chile
              \email{paula.jofre@mail.udp.cl}
         \and
             Department of Electrical Engineering and Computer Science, 
Massachusetts Institute of Technology,
50 Vassar Street, Cambridge, MA 02139, USA}

   \date{Received November 20, 2019; accepted -- }

% \abstract{}{}{}{}{} 
% 5 {} token are mandatory
 
  \abstract
   {The physical processes driving chemical evolution in the Milky Way can be probed using the distribution of abundances in low-mass FGK type stars in space phase at different times.  { During their final stages of evolution stars experience nucleosynthesis several times, each at different timescales and producing different chemical elements, so finding abundance ratios that have simple variations across cosmic times remains a challenge.} Using the sample of 80 solar twins for which ages and abundances of 30 elements were measured with high precision by \citet{Bedell_2018ApJ...865...68B}, we searched for all possible abundance ratios combinations that show linear trends with age. We found 55 such ratios, all combining a \nc\ element and another one produced by different nucleosynthesis channels. We recovered the ratios of [Y/Mg], [Ba/Mg] and [Al/{  Y}] reported previously in the literature, and found that [C/Ba] had the largest dependency with age with a slope of $0.049 \pm 0.003~\mathrm{dex/Gyr}$, imposing constraints on the {  magnitude of time dependency of abundance ratios in solar twins. Our results suggest using \spro\ elements, in lieu of Fe, as a reference for constraining chemical evolution models of the solar neighbourhood. } Our study illustrates hows a large variety of chemical elements measured from high resolution spectra is key in facing current challenges in understanding the formation and evolution of our Galaxy.  }


   \maketitle
%
%-------------------------------------------------------------------

\section{Introduction} \label{sec:intro}

Chemical abundances of {  FGK low-mass} stars have shown to be powerful tracers of the Milky Way evolution. This is because we assume the chemical composition of these stars' birthplaces is recorded in their { photosphere}. Since chemical composition evolves from one stellar generation to the next, when we combine this information with stellar age and phase-space, it is possible to reconstruct the different physical processes that shaped the Milky Way across cosmic times \citep{Freeman_2002ARA&A..40..487F}. 

The way chemistry in the Milky Way evolves is mainly due to stellar nucleosynthesis \citep{Nomoto_2013ARA&A..51..457N, Karakas_2016ApJ...825...26K}.  Newly synthesized metals are sent back to the interstellar medium and are recycled in new stars, and thereby their elemental abundances increase with time \citep{daSilva_2012A&A...542A..84D}. It is thus {  commonly assumed that}   the chemical composition of stars {  imprinted in} their atmospheres remains unchanged in stellar lifetime, but it evolves in time through inheritance between  stellar generations \citep{Jofre_2017MNRAS.467.1140J}. Hence, each chemical abundance  measured from a stellar spectrum can be understood as a trait for {  reconstructing  the chemical evolution of the Galaxy. }

{  Of course, not all chemical elements are informative traits for chemical evolution studies, because in reality some of them may change through stellar lifetime. There} are inner processes in stellar evolution such  diffusion \citep{Dotter_2017ApJ...840...99D} or planet formation/engulfment \citep{Melendez_2009ApJ...704L..66M, Tucci_2019A&A...628A.126M} that alter the chemical composition in stars.  {  Other inner mixing processes produced during dredge-up or induced thanks to rotation can alter abundances such as [C/N] or Li \citep{Masseron_2017MNRAS.464.3021M, Aguilera_2018A&A...614A..55A}. Nowadays we have to assume that a selection of chemical abundances remain approximately intact in FGK main-sequence stars and can be used to trace chemical evolution. }    

Few years ago, \citet{daSilva_2012A&A...542A..84D}  studied several traits for chemical evolution studies of solar-type stars. In particular, $\mathrm{[Y/Mg]}$, $\mathrm{[Sr/Mg]}$, $\mathrm{[Y/Zn]}$ and $\mathrm{[Sr/Zn]}$ showed a significant trend as a function of age for their entire sample. When performing a linear fit of $\mathrm{[Y/Mg]}$  as a function of stellar age for a sample of nearby solar twins, \citet{Nissen_2015A&A...579A..52N} obtained slope of $ \sim 0.04~\mathrm{dex/Gyr}$  and interpreted this relation using nucleosynthesis arguments, namely that massive stars produced $\mathrm{Mg}$ and expelled it into the interstellar medium at a different timescale than intermediate-mass reaching the AGB and producing $\mathrm{Y}$.   This has motivated discussions about the possibility of using abundance ratios as ``chemical clocks'', in particular if ages are difficult to measure.  The exact dependency of  $\mathrm{[Y/Mg]}$ {   or $\mathrm{[Ba/Mg]}$} with age is still debated as it seems to further depend on stellar properties such as metallicity \citep{Feltzing_2017MNRAS.465L.109F, Skuladottir_2019arXiv190810729S} but not evolutionary state, at least before the AGB phase  \citep{Slumstrup_2017A&A...604L...8S}.  In addition to  $\mathrm{[Y/Mg]}$ and  $\mathrm{[Ba/Mg]}$, $\mathrm{[Al/Y]}$ also have shown similar dependencies with ages \citep{Nissen_2016A&A...593A..65N, Spina_2016A&A...593A.125S},  at least for solar twins. 

Several other works that benefit from accurate age estimates have studied trends of abundance ratios as a function of age \citep[][to mention recent works]{Bedell_2018ApJ...865...68B, Spina_2018MNRAS.474.2580S, delgado_2019A&A...624A..78D, feuillet_2018MNRAS.477.2326F}. Most of these analyses have been restricted to using abundance ratios in their classical form, namely as a function of iron.  {  Motivated by the fact that the ``chemical clocks'' discussed above do not include iron, we searched for other combinations in a sample of 30 elemental abundances. We found 55 relations that are as significant as those previously mentioned.   This new set of ratios can be used} for the same applications as $\mathrm{[Y/Mg]}$.  In addition, they can be applied to models of Galaxy evolution and used to build stellar phylogenies \citep[][Jackson et al. in prep]{Jofre_2017MNRAS.467.1140J}. In this letter we report these findings. 

\section{Data and Methods}
We considered the sample of 80 solar twins (including the Sun) analysed by \citet[and references therein]{Bedell_2018ApJ...865...68B} and \citet{Spina_2018MNRAS.474.2580S}. This is the largest sample of Solar Twins analysed using the differential technique for high-precision estimates of the {  abundances of 30 elements as well as stellar ages.  The ages and abundance data, as well as the list of stars, can be found in these papers referenced. }

In short, {  \cite{Bedell_2018ApJ...865...68B} and \cite{Spina_2018MNRAS.474.2580S} measured chemical abundances} using HARPS@La Silla spectra taken from the public ESO Science Archive Facility. The spectra have very high resolving power (100,000). For multiple observations, the spectra were stacked to achieve very high signal-to-noise of above 200 pix$^{-1}$. In addition, spectra taken with the MIKE@Las Campanas of slightly lower resolution (65,000 to 80,000) but equal SNR  were used to complement the wavelength ranges not covered by HARPS. This allowed  {  the researchers} to measure abundances of elements like oxygen from the O triplet at 770 nm. Stellar parameters and abundances were derived using 1D LTE analysis based on equivalent widths. Stellar ages were estimated using fits to isochrones following the same procedure as described in \citet{Spina_2018MNRAS.474.2580S}. 

We considered all possible combinations of abundance ratios of all 30 elements   and performed a linear regression fit to these ratios as a function of age.  Unlike \citet{Bedell_2018ApJ...865...68B} and \cite{Spina_2018MNRAS.474.2580S}, we considered all stars, including the very old ones that are suspected to be part of the thick disk {  and could follow a different chemical enrichment history}. The regression fit was done using the  python 2.7 routine {\tt stats.linregress} from the {\tt scipy} library.

\begin{figure*}[htbp]
\hspace{-0.5in}
\includegraphics[scale=0.8]{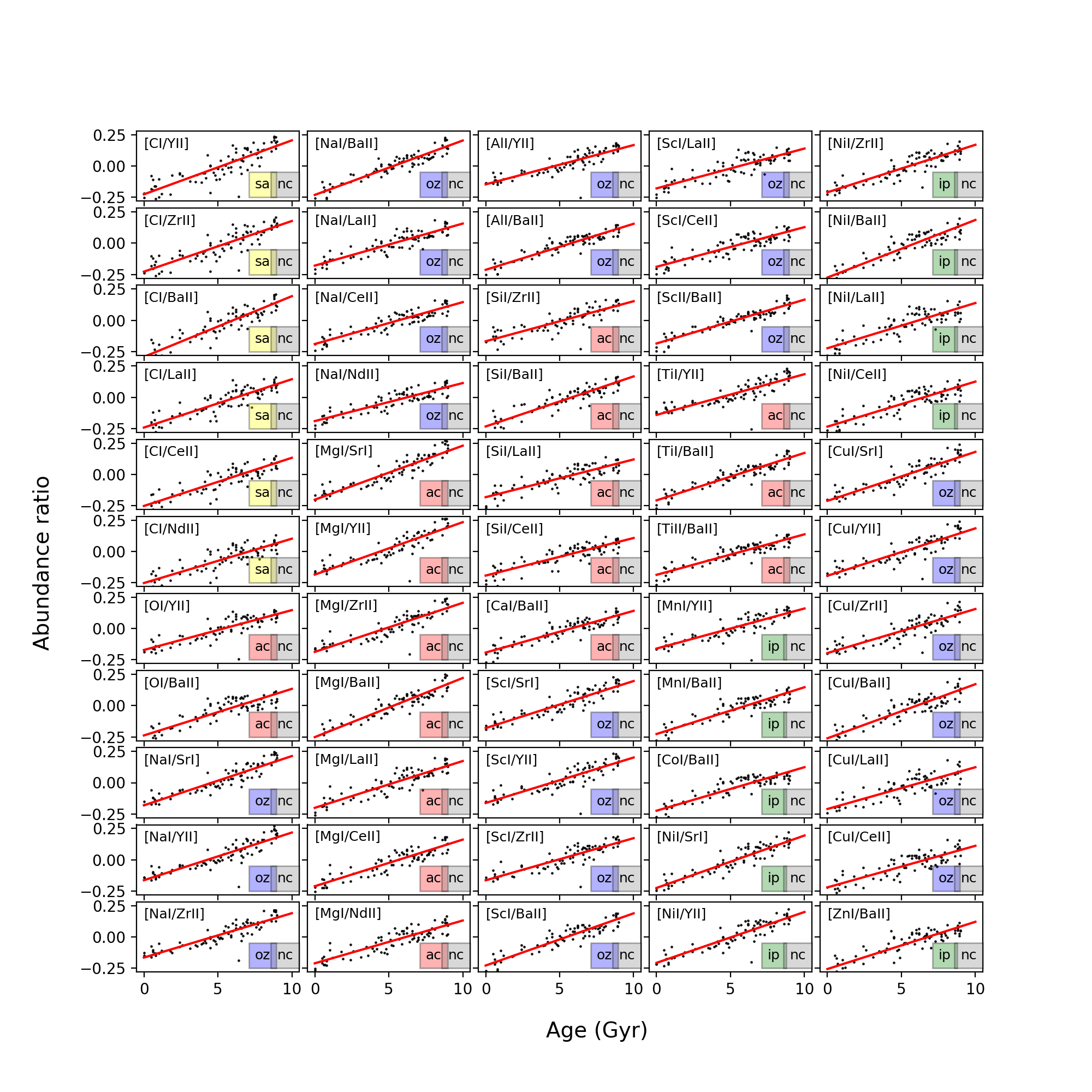}
\caption{Selected abundance ratios as a function of age in solar twins together with the linear fits of the trends. The abundance ratio is indicated in each panel at the top left, while at the bottom right the nucleosynthetic family of each abundance in a different colour (see Table~\ref{tab:families} for the definition of each label).    }
\label{fig:traits}
\end{figure*}

 Only the traits whose linear fit had an absolute slope greater than $0.03 ~ \mathrm{dex}/\mathrm{Gyr}$ and a correlation factor greater than 0.7 were considered. Choosing that threshold for the slope is motivated by the normally lower than $0.03 ~ \mathrm{dex}/\mathrm{Gyr}$ slope seen in the linear fits {  of abundance ratios as a function of Fe by} \citet{Bedell_2018ApJ...865...68B}. We were interested in finding relations that are as significant as the [Y/Mg] or [Ba/Mg] relations discussed in the literature, which have a slope greater than $0.03 ~ \mathrm{dex}/\mathrm{Gyr}$. We note that different fitting techniques, selection of stars, thresholds for the slopes and correlation coefficients might yield slightly different results and therefore a different final sample of selected traits. {  Therefore, the results of the linear fits of all abundance combinations as a function of age can be found in the online material. } %However, in this letter our goal is to report the advantages of considering abundance ratios of elements different to iron as traits. 
 
\begin{table*}[t]
\caption{Linear regression fit coefficients. Error:standard error ($\times 10^2$); $r^2$: correlation coefficient}
\begin{center}
\begin{small}
\begin{tabular}{c|cccc ||| c|cccc}
\hline
Trait & slope & intercept & error & $r^2$ & Trait & slope & intercept & error & $r^2$ \\
 & (dex/Gyr) & dex & (dex/Gyr) &  & & (dex/Gyr) & dex & (dex/Gyr) &  \\
\hline
$\mathrm{[CI/YII]}$ & 0.043 & -0.226 & 0.304& 0.721 &&&\\
$\mathrm{[CI/ZrII]}$ & 0.040 & -0.228 & 0.277& 0.733 &
$\mathrm{[CI/BaII]}$ & 0.049 & -0.293 & 0.255& 0.822 \\
$\mathrm{[CI/LaII]}$ & 0.039 & -0.242 & 0.245& 0.762 &
$\mathrm{[CI/CeII]}$ & 0.038 & -0.252 & 0.234& 0.777 \\
$\mathrm{[CI/NdII]}$ & 0.035 & -0.252 & 0.248& 0.724 &
$\mathrm{[OI/YII]}$ & 0.032 & -0.171 & 0.229& 0.713 \\
$\mathrm{[OI/BaII]}$ & 0.037 & -0.238 & 0.240& 0.754 &
$\mathrm{[NaI/SrI]}$ & 0.039 & -0.181 & 0.267& 0.736 \\
$\mathrm{[NaI/YII]}$ & 0.038 & -0.164 & 0.239& 0.765 &
$\mathrm{[NaI/ZrII]}$ & 0.035 & -0.165 & 0.212& 0.783 \\
$\mathrm{[NaI/BaII]}$ & 0.043 & -0.231 & 0.187& 0.874 &
$\mathrm{[NaI/LaII]}$ & 0.034 & -0.180 & 0.195& 0.792 \\
$\mathrm{[NaI/CeII]}$ & 0.033 & -0.190 & 0.177& 0.820 &
$\mathrm{[NaI/NdII]}$ & 0.030 & -0.190 & 0.185& 0.775 \\
$\mathrm{[MgI/SrI]}$ & 0.043 & -0.202 & 0.281& 0.752 &
$\mathrm{[MgI/YII]}$ & 0.042 & -0.185 & 0.255& 0.776 \\
$\mathrm{[MgI/ZrII]}$ & 0.039 & -0.187 & 0.232& 0.786 &
$\mathrm{[MgI/BaII]}$ & 0.047 & -0.252 & 0.201& 0.876 \\
$\mathrm{[MgI/LaII]}$ & 0.037 & -0.201 & 0.221& 0.786 &
$\mathrm{[MgI/CeII]}$ & 0.037 & -0.211 & 0.204& 0.810 \\
$\mathrm{[MgI/NdII]}$ & 0.034 & -0.212 & 0.212& 0.769 &
$\mathrm{[AlI/YII]}$ & 0.031 & -0.146 & 0.219& 0.725 \\
$\mathrm{[AlI/BaII]}$ & 0.037 & -0.213 & 0.172& 0.854 &
$\mathrm{[SiI/ZrII]}$ & 0.032 & -0.168 & 0.235& 0.703 \\
$\mathrm{[SiI/BaII]}$ & 0.040 & -0.233 & 0.219& 0.811 &
$\mathrm{[SiI/LaII]}$ & 0.030 & -0.182 & 0.211& 0.724 \\
$\mathrm{[SiI/CeII]}$ & 0.030 & -0.193 & 0.205& 0.734 &
$\mathrm{[CaI/BaII]}$ & 0.033 & -0.192 & 0.179& 0.817 \\
$\mathrm{[ScI/SrI]}$ & 0.038 & -0.180 & 0.274& 0.707 &
$\mathrm{[ScI/YII]}$ & 0.036 & -0.162 & 0.246& 0.737 \\
$\mathrm{[ScI/ZrII]}$ & 0.034 & -0.164 & 0.223& 0.745 &
$\mathrm{[ScI/BaII]}$ & 0.042 & -0.229 & 0.204& 0.843 \\
$\mathrm{[ScI/LaII]}$ & 0.032 & -0.178 & 0.221& 0.727 &
$\mathrm{[ScI/CeII]}$ & 0.032 & -0.189 & 0.200& 0.763 \\
$\mathrm{[ScII/BaII]}$ & 0.035 & -0.186 & 0.153& 0.870 &
$\mathrm{[TiI/YII]}$ & 0.033 & -0.141 & 0.238& 0.705 \\
$\mathrm{[TiI/BaII]}$ & 0.038 & -0.207 & 0.169& 0.865 &
$\mathrm{[TiII/BaII]}$ & 0.032 & -0.187 & 0.163& 0.834 \\
$\mathrm{[MnI/YII]}$ & 0.032 & -0.160 & 0.226& 0.720 &
$\mathrm{[MnI/BaII]}$ & 0.037 & -0.227 & 0.204& 0.812 \\
$\mathrm{[CoI/BaII]}$ & 0.035 & -0.224 & 0.208& 0.781 &
$\mathrm{[NiI/SrI]}$ & 0.042 & -0.225 & 0.253& 0.779 \\
$\mathrm{[NiI/YII]}$ & 0.041 & -0.208 & 0.226& 0.805 &
$\mathrm{[NiI/ZrII]}$ & 0.038 & -0.210 & 0.221& 0.791 \\
$\mathrm{[NiI/BaII]}$ & 0.046 & -0.275 & 0.234& 0.832 &
$\mathrm{[NiI/LaII]}$ & 0.036 & -0.224 & 0.251& 0.727 \\
$\mathrm{[NiI/CeII]}$ & 0.036 & -0.234 & 0.238& 0.745 &
$\mathrm{[CuI/SrI]}$ & 0.039 & -0.211 & 0.270& 0.729 \\
$\mathrm{[CuI/YII]}$ & 0.038 & -0.194 & 0.243& 0.757 &
$\mathrm{[CuI/ZrII]}$ & 0.035 & -0.196 & 0.236& 0.741 \\
$\mathrm{[CuI/BaII]}$ & 0.043 & -0.261 & 0.223& 0.828 &
$\mathrm{[CuI/LaII]}$ & 0.033 & -0.210 & 0.240& 0.712 \\
$\mathrm{[CuI/CeII]}$ & 0.033 & -0.221 & 0.234& 0.720 &
$\mathrm{[ZnI/BaII]}$ & 0.038 & -0.257 & 0.231& 0.773 \\
\hline
\end{tabular}
\end{small}
\end{center}
\label{tab:traits}
\end{table*}%

\begin{table}[t]
\caption{Nucleosynthetic families of elements considered in this study. The label is used as reference in Figure~\ref{fig:traits}. }
\begin{center}
\begin{tabular}{c|c|c}
\hline
family & label & elements \\
\hline
Carbon & sa & C\\
$\alpha-$capture & {\tt ac} & O, Mg, Si, S, Ca, Ti \\
odd$-Z$ &{\tt oz} & Na, Al, Cu, Sc, V\\
iron$-$peak & {\tt ip} & Cr, Mn, Fe, Co, Ni, Zn\\
$n-$capture & {\tt nc} & Sr, Y, Zr, Ba, La, Ce\\
&& Pr, Nd, Sm, Eu, Gd, Dy\\ 
\hline
\end{tabular}
\end{center}
\label{tab:families}
\end{table}%

\section{Results}

 The linear correlation coefficients for the selected traits are listed in Table~\ref{tab:traits}. %, where we indicate the value of the slope and intercept of the linear fit. We also indicate the standard error of the fit (multiplied by 100) and the correlation factor ($r^2$).  
 The fit is further plotted in Figure~\ref{fig:traits} alongside with the abundance ratio for each star as a function of age.  Each trait is plotted in a different panel, and is indicated in the top left side.  %For reference, all elements considered in this analysis sorted by their nucleosynthetic family (see details below), are listed in Table~\ref{tab:families}.
 
 For better interpretation of the results, at the bottom right side of  each panel we have added two boxes with labels and colours, which help to attribute the chemical element involved in the trait to a nucleosynthesis family. The left-hand side box represents the family of the element in the numerator while the right-hand side box represents the family of the element in the denominator of the abundance ratio. The families were chosen following the classification indicated in Table~\ref{tab:families}. This classification comes from discussions in \cite{Nomoto_2013ARA&A..51..457N}, except the neutron-capture family, for which we simply considered the elements analysed by \citet{Spina_2018MNRAS.474.2580S}. As extensively discussed in that work \citep[see also][]{Bisterzo_2014ApJ...787...10B, Karakas_2016ApJ...825...26K, Battistini_2016A&A...586A..49B}, it is difficult to disentangle how much of these elements is produced by the slow neutron-capture process ($s$) and how much by the rapid neutron-capture process ($r$), each having a different site. This is why we decided to include all elements in the same family here.

We note  that \cite{Nomoto_2013ARA&A..51..457N} did not consider titanium as an $\alpha-$capture element, since this element's theoretical prediction does not match the observations \citep{kobayashi_2006ApJ...653.1145K}. However, observationally Ti behaves like an $\alpha-$element, at least in terms of the abundance trends as a function of $\mathrm{[Fe/H]}$, which is why we include it in this family.  Carbon belongs to {  a special family that we call {\tt sa}, because carbon has multiple unique formation sites.  It can be produced in AGB stars, via $s$-process, or from massive stars, via $\alpha-$capture \citep[see][for extensive discussions]{Amarsi_2019A&A...630A.104A}. While other elements are also produced by a variety of channels,  carbon's mass production is believed to be significant in all these channels.} %In Figure~\ref{fig:traits} we have assigned to each family box a colour, yellow for carbon, red for  \ac, blue for odd$-Z$, green for \ip\ and grey for \nc. 

%\subsection{Different nucleosynthesis channels}
We find that all selected traits involve different families. %On the contrary, traits with elements of same families do not vary significant with age, enforcing the idea that their production site is similar. 
{  The ratios selected here consider elements in their [X/Fe] form with opposite (increasing vs. decreasing) trends with age. In fact,  \cite{Spina_2018MNRAS.474.2580S} obtained negative trends of the \nc\ elements in their [X/Fe] form as a function of age, and \cite{Bedell_2018ApJ...865...68B} obtained both positive and negative slopes for the other elements. The traits found here correspond to abundance ratios of elements that showed a general positive trend in their [X/Fe] vs. age correlation in \cite{Bedell_2018ApJ...865...68B}}. % We comment that some abundance ratios might have a dependency with age, but it might be smaller than the cut of $0.03~\mathrm{dex/Gyr}$ imposed for our selection of traits. %While each element within a given family might have a different nucleosynthetic site, overall the dependency with time of the production of both elements is similar (at least to the level of a change that is below $0.03~ \mathrm{dex/Gyr}$). 
%If we look for maximise change of abundance with age, it is best to consider elements produced by different channels.  

We further find  that the denominator of our selected abundance ratios is always a \nc\ element, including the already  studied traits [Mg/Y] and [Mg/Ba]. The slopes in our fits are 0.042 and 0.047 $\mathrm{dex/Gyr}$, respectively. This agrees well with previous findings \citep{Nissen_2015A&A...579A..52N, Tucci_2016A&A...590A..32T}.    The \nc\ elements selected here are known to have {  more than 50\% contribution from the  \spro\ in the Sun \citep[see][for a discussion on this sample]{Spina_2018MNRAS.474.2580S}. The percentages were calculated by \cite{Bisterzo_2014ApJ...787...10B}.} Elements with lower {  percentage in \spro\ contribution for the solar abundances }(like Eu, Sm, or Dy) were not selected to vary enough with age, according the restrictions imposed in this analysis. The trend of [C/Ba] with age has the largest slope in our sample of $0.049 \pm 0.003~ \mathrm{dex}/\mathrm{Gyr}$.  {  This puts a constraint on the maximum change of abundance ratio with age for this sample of solar metallicity stars.} 

The final 55 abundance ratios contain 6 with carbon, 17 with \ac\ elements, 10 with \ip\ elements and 22 with \oddz\ elements. Among the \ac, we obtain traits including O, Mg, Si and Ti.  The element S was not selected.  Among the \ip, we obtain traits including Mn, Co, Ni and Zn. It is interesting that Fe was not selected, since it is the standard element used for abundance ratios. %We note however this might be a selection effect, since the sample selected of solar twins did not show and correlation of Fe with age.  
Cr was also not selected. 
Regarding the \oddz,  we find Na, Al, Sc and Cu selected, but not V. 

\begin{table}[t]
\caption{ { Mean slopes and their standard deviation for traits separated by \nc\ element or by family. The number column indicates the times this particular element/family was selected}}
\begin{center}
\begin{tabular}{c|ccc}

\hline
Element & number & mean slope & std. dev\\
family & & dex/Gyr & dex/Gyr \\
\hline
 Ba & 16 & 0.039 & 0.005 \\
 Ce &  7 & 0.034 & 0.003 \\
 La &  7 & 0.034 & 0.003 \\
 Nd &  3 & 0.033 & 0.002 \\
 Sr &  5 & 0.040 & 0.002 \\
 Y & 10 & 0.037 & 0.004 \\
 Zr &  7 & 0.036 & 0.003 \\
\hline
 {\tt ac} & 17 & 0.036 & 0.005 \\
 {\tt ip} & 10 & 0.038 & 0.004 \\
 {\tt oz} & 22 & 0.036 & 0.004 \\
 {\tt sa} &  6 & 0.041 & 0.004 \\
\hline
\end{tabular}
\end{center}
\label{tab:stats}
\end{table}%

{  In order to investigate the abundance ratios further, we grouped the traits by the different \nc\ elements  or by the families of the numerator of the abundance ratios. The mean and standard deviation of the slopes for each of the groups can be found in Table~\ref{tab:stats}, where we also indicate the total number a given abundance/family was selected.

\begin{figure}[htbp]
\begin{center}
\includegraphics[scale=0.6]{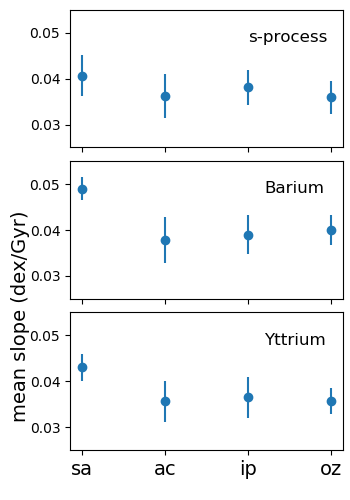}
\caption{Mean and standard deviations of slopes grouped by their families. The first panel corresponds to all selected \nc\ elements while the second and third panel consider only the results for Ba and Y, respectively.  }
\label{fig:slopes}
\end{center}

\end{figure}

Regarding the \nc\ elements, the largest mean slopes are obtained for Sr and Ba, although the mean slope for Ba has the largest standard deviation. However, there is no relation between the the solar \spro\ percentage contribution and the slopes for the age trends, if we consider the values adopted by \cite{Spina_2018MNRAS.474.2580S} which are obtained by \cite{Bisterzo_2014ApJ...787...10B}. 

When looking at the mean slopes of the different families,  carbon is the one with the largest mean slope, but they all seem to be of comparable magnitudes, especially when considering their standard deviations.  These results can be further visualised  in Figure~\ref{fig:slopes}, where we plot the mean slope and standard deviation of the mean for each family. The first panel contains all selected \nc\ elements, while the second and third panels show the results for Ba and Y, because these were shown to contribute to ratios of elements in all families. We note that for Ba and Y, the error bar of the {\tt sa} family corresponds to the uncertainty of the linear fit (see Table~\ref{tab:traits} for [C/Ba] and [C/Y], respectively). All family combinations behave similarly, with C systematically yielding higher slopes. 

}

%{  Calculating the mean of the slopes in the regression fit  (the values indicated in Table~\ref{tab:traits}) for each family finding that there is no significant difference among the groups. For carbon the mean slope is of $ 0.04078 ~\mathrm{dex / Gyr}$, for the \ac\ family of $0.0365~ \mathrm{dex / Gyr}$, for the \ip\ family of $ 0.0381~ \mathrm{dex / Gyr}$ while for the \oddz\ family the slope is of $0.03601 ~\mathrm{dex / Gyr}$.  While carbon shows a slightly larger slope, we note that this number was calculated with only 6 traits. In general, all families have similar trends as a function of age - ADD changing \nc\ elements} 

\section{Discussion}

Neutron-capture elements, in particular those produced by the \spro, have shown to add crucial information in chemical evolution studies \citep[see recent study of ][]{Skuladottir_2019arXiv190810729S}. Their gradual and slow production as intermediate-mass stars reach the AGB phase and produce stellar winds carrying the newly synthesised elements into the ISM.  Hence, new generations of stars formed within this environment will inherit a gradual increase of \spro\ elements. Since the production of \spro\ depends on the overall metallicity of the progenitor star \citep{Karakas_2016ApJ...825...26K}, the enrichment of \spro\ elements is very dependent on the local conditions and star formation history of the previous generations.  

With this in mind, the relation of [Mg/Y] or [Mg/Ba] as a function of age has been explained with chemical evolution principle in the literature.% Because Mg is an \ac\ element produced primarily in massive stars,  large amounts with respect to other elements are released early on in Galaxy formation, before the enrichment due to lower mass stars such as \spro\ could take place.  
It is thus expected that not only would these 2 ratios show significant trends as a function of stellar age, but other ratios involving both \ac\ and \spro\ elements, such as the other 15 cases resulting from our exercise, would as well. 
Although C behaves as both \ac\ and \nc\ element, the fact we find 6 traits with strong dependency of C over a \spro\ element with age suggests that the C contribution to chemical evolution from \spro\ is lower than from \ac, at least at solar metallicities.   {  This reasoning might contradict the results of \cite{Amarsi_2019A&A...630A.104A} who studied the [C/O] ratio for different stellar populations. The fact that this ratio is negative for halo stars and zero for disk stars, suggests that the carbon's  \spro\ contribution plays an important role as metallicity increases. That conflicts with our reasoning that slopes in trends of abundance ratios with age increase if production sites for the elements differ. Our sample however contains stars of same metallicity but different ages. \cite{Limongi_2018ApJS..237...13L} presented an extensive study of nucleosynthesis of massive stars obtaining that the relative production of C with respect to O increases with metallicity.  The large slopes of C with respect to several \spro\ elements found here offer new alternatives to study the production mechanisms of this element. }

The relation of the \oddz\ element Al with respect to Y has also caught the attention in the literature \citep{Nissen_2016A&A...593A..65N}. It is explained following similar nucleosynthetic arguments as for [Mg/Y], that the rapid production of Al in core-collapse SNe of massive stars is compensated with the increasing contribution of Y from lower-mass AGB stars in time. Na is produced in a similar way as Al, being alternatives to stellar population studies \citep{Hawkins_2015MNRAS.453..758H}. In \citet{Nissen_2016A&A...593A..65N}  it is further discussed how $\mathrm{[Cu/Na]}$ remains constant over time, suggesting a similar production site for these elements.  Our traits for the \oddz\ family contain both Cu and Na elements with respect to similar \nc\ elements.

%Because the \spro\ production is dependant on metallicity, for these solar twins, which already have experienced metal enrichment due to Type Ia  supernovae, the site for more \spro\ production is available. It is thus possible to understand why the increasing contribution of iron-peak due to these supernovae will imply a subsequent increase of production of \spro\ elements. 
{  Ni seems to be the \ip\ element with the strongest influence, which could be explained by the strong [Na/Ni] correlation found for these type of stars  \citep{Nissen_2015A&A...579A..52N} and the strong positive [Ni/Fe] correlation found by \cite{Bedell_2018ApJ...865...68B}. It is interesting to note that while we set the threshold for selecting our traits above the typical slopes found for [X/Fe], we do select \ip\ elements. This means that \ip\ family does not necessarily have a weaker dependency with age when considering its ratio with \spro\ abundances (see Table~\ref{tab:stats}).  The \ip\ family has indeed a variety of mechanisms to create elements, and is expelled to the interstellar medium by different supernovae. [Mn/Fe], for example, has significant trends as a function of [Fe/H], and Zn is also greatly produced by very massive stars \citep{kobayashi_2006ApJ...653.1145K}.  }

%It is worth to comment that the traits found here are for solar twins, which means that even every star formed from gas that had experienced extensive chemical enrichment until solar metallicities. Other traits might be more relevant at lower metallicities. Nonetheless, this study shows the importance of considering a large variety of chemical abundances and how they can provide different insights in chemical evolution of the Milky Way.  

\section{Conclusion}
We present the results of abundance ratios as a function of stellar ages for a set of 80 solar twins for which 30 different chemical elements were measured with high precision by \citet[][and references therein]{Bedell_2018ApJ...865...68B}. By selecting only the trends whose linear fit yielded a correlation factor above 70\% and a slope above $0.03~\mathrm{dex/Gyr}$ we found 55 abundance ratios, including the widely discussed [Mg/Y] and [Mg/Ba]  ``chemical clocks''.  

All of them included two different nucleosynthetic families, and  all included one \nc\ element whose \spro\ percentage to the solar abundance is larger than the \rpro\ one.  The maximum slope found was for $\mathrm{[C/Ba]}$ of $\sim0.05  \pm 0.003 ~ \mathrm{dex}/\mathrm{Gyr}$.  There seems to be no abundance ratio evolving faster than $\sim0.05 ~ \mathrm{dex}/\mathrm{Gyr}$ in solar twins.  

The new era of industrial stellar abundances from on-going and future spectroscopic surveys has just begun, and is already providing the community with datasets of chemical abundances of elements in all families considered here \citep{Jofre_2019ARA&A..57..571J}. In particular, thanks to Gaia and seismic missions such as  {\it Kepler} or TESS, it is becoming possible to estimate age distributions of many stars in the Milky Way. Investigating how the traits for chemical evolution found here behave in different conditions can provide new insights for constraining the chemodynamical models of the formation and evolution of our Galaxy.  Much of this information is already available, we should take the opportunity to explore it by  considering abundance ratios as a function the \spro\ elements.  
\begin{acknowledgements}
We thank the referee for their positive and thoughtful comments, which helped to significantly order the discussions. We warmly acknlowledge K. Yaxley, D. Boubert and P. Das for lively discussions on evolution of traits in general. We further acknowledge support from the MIT International Science and Technology Initiatives (MISTI) grant, which funded H.J. research visit from which these results were obtained. P.J. is partially funded by FONDECYT Iniciaci\'on grant Number 11170174.   M.T. and P.J. thank the Joint Committee ESO-Chile for postdoctoral financial support, and ECOS-Conicyt for enabling travel and relevant discussions that lead to this paper. 
\end{acknowledgements}

\bibliography{references}{}
\bibliographystyle{aasjournal}

\end{document}